\documentclass{elsart}
\usepackage{amsmath,amssymb,graphicx}

\begin{document}

\begin{frontmatter}

\title{Renormalization Group Equation for Low Momentum Effective
Nuclear Interactions}
\author{S.K. Bogner, A. Schwenk, T.T.S. Kuo, and G.E. Brown\thanksref{EMail}}
\address{Department of Physics and Astronomy, State University of
New York,\\Stony Brook, NY 11794-3800}
\thanks[EMail]{E-mail: bogner,kuo,aschwenk,popenoe@nuclear.physics.sunysb.edu}

\begin{abstract}
We consider two nonperturbative methods originally used to derive
shell model effective interactions in nuclei. These methods have
been applied to the two nucleon sector to obtain an
energy independent effective interaction $V_{\text{low k}}$,
which preserves the low momentum half-on-shell $T$ matrix and the
deuteron pole, with a sharp cutoff imposed on all intermediate state
momenta. We show that $V_{\text{low k}}$ scales with
the cutoff precisely as one expects from renormalization group
arguments. This result is a step towards reformulating traditional
model space many-body calculations in the language of effective field
theories and the renormalization group. The numerical 
scaling properties of $V_{\text{low k}}$ are observed to be in 
excellent agreement with our exact renormalization group equation.

\vspace{0.5cm}

\noindent{\it PACS:}
21.30.Fe;          
13.75.Cs;	   
11.10.Hi;          
03.65.Nk \\	   
\noindent{\it Keywords:} Nucleon-Nucleon Interactions; Effective
Interactions; Renormalization Group; Scattering Theory
\end{abstract}
\end{frontmatter}

\section{Introduction}

There has been much work over the past decade applying
the techniques of effective field theory (EFT) and the
renormalization group (RG) to low energy nuclear systems such as
the nucleon-nucleon force, finite nuclei, and nuclear matter~\cite{encyc}.
Conventional nuclear force models such as the Paris, Bonn, and Argonne 
potentials incorporate the same asymptotic tail generated
by one pion exchange, as the long wavelength structure of the
interaction is unambiguously resolved from fits to low energy phase shifts and
deuteron properties. The short wavelength part of the interaction is then 
generated by assuming a specific dynamical model based on heavy meson 
exchanges, combined with phenomenological treatments at very small distances. 
Such approaches are necessarily model dependent, as the low energy 
two-nucleon properties are insufficient to resolve the short distance 
structure. Such model dependence often appears in many-body 
calculations, e.g. the Coester band in nuclear matter, 
when highly virtual nucleons probe the short distance 
structure of the interaction. The EFT approach eliminates the 
unsatisfactory model dependence of conventional    
force models and provides an effective description  
that is consistent with the low energy symmetries of the
underlying strong interactions (QCD). This is accomplished by keeping only   
nucleons and pions as explicit degrees of freedom, as dictated by  
the spontaneously broken chiral symmetry of QCD.   
All other heavy mesons and nucleon resonances  
are considered to be integrated out of the theory,
their effects contained inside the renormalized pion exchange and
scale dependent coupling constants that multiply model independent
delta functions and their derivatives~\cite{lepage}. 
No underlying dynamics are assumed for the heavy mesons and nucleon 
resonances, as they simply cannot be resolved from low energy data.
Since RG decimations generate all possible
interactions consistent with the symmetries of the underlying
theory, it is sufficient to consider all interactions
mandated by chiral symmetry and then tune the couplings to the low energy
data. Power counting arguments are then used to truncate the
number of couplings that need to be fit to experiment, thus 
endowing the EFT with predictive power and the ability to  
estimate errors resulting from the truncation.
Moreover, the breakdown of the EFT at a given scale signals 
the presence of new relevant degrees of freedom that must be considered
explicitly to properly describe phenomena at that scale. 

Similar concepts of integrating out the high energy modes have long been 
used to derive effective interactions in nuclei within a truncated
model space, e.g. the sd-shell for the two valence nucleons in
$^{18}$O. Starting from the vacuum two-body force in the full
many-body Hilbert space, one can construct an effective theory for the
low-lying excitations provided the effective interaction encodes the 
effects of the integrated high energy modes. Although the
traditional model space methods apparently share similarities to the
modern RG-EFT approaches, these have not been exploited
until recently~\cite{Vlowk,schematic,Haxton1,Haxton2} 
and not sufficiently in any realistic nuclear
many-body calculation. In the traditional approaches, there has
been little success in predicting how the effective interaction
changes with the model space size. In RG language, no beta
functions have been derived that would allow one to calculate an
effective interaction in one convenient choice of model space, and
then evolve the effective theory to any other scale by following
the flow of the beta function. Moreover, one could push
the analogy with EFT further by projecting the effective
interaction onto a few leading operators, with the ability to
reliably estimate the errors on calculated nuclear spectra 
resulting from this truncation. Therefore, it is of greatest 
interest to address the issue of calculating beta functions 
within the framework of model space effective interaction methods 
and to exploit these powerful similarities with the RG-EFT approach. 

Two well known methods for deriving {\it energy independent} 
model space interactions are the Kuo-Lee-Ratcliff
(KLR) folded diagram theory~\cite{KLR} and the related similarity
transformation method of Lee and Suzuki (LS)~\cite{LS}. 
The authors have applied these methods to the nucleon-nucleon problem in vacuum
where the model space was taken to be plane wave states with
relative momenta $k < \Lambda$. The resulting unique low momentum
potential $V_{\text{low k}}$ preserves the deuteron binding energy
and the low energy half-on-shell $T$ matrix, but with all
intermediate state summations cut off at $\Lambda$~\cite{Vlowk}.
In this paper, we restrict our analysis to the two-body problem in
free space. We show that the model space interaction
$V_{\text{low k}}$ scales with $\Lambda$ in the same
way one would expect from a exact RG treatment of the scattering
problem. In this way, we show that the methods originally used to
derive model space interactions in nuclei can be interpreted in
modern language as renormalization group decimations, at least for
two-body correlations in the nucleus. This work is a step towards
reformulating traditional nuclear many-body methods in a manner
consistent with the more systematic and controlled RG-EFT
approaches. To the best of our knowledge, this is a genuinely new
result as previous RG studies have dealt with energy dependent
effective potentials~\cite{RG2body}. It is also the first RG flow
study of realistic nucleon-nucleon interactions. From a practical perspective,
it is simpler to use energy independent effective interactions in
many-body calculations, as one does not have to recalculate the
interaction vertex depending on the energy variable of the
propagator it is linked to in a particular diagram.

\section{RG from half-on-shell $T$ Matrix Equivalence}

Let us begin with a RG treatment of the scattering problem.
Working in a given partial wave we denote the bare (Bonn, Paris,
Argonne, etc.) two-body potential as $V_{\text{NN}}$, although the
following developments are general and apply to any
nonrelativistic scattering problem. Employing principle value
propagators we have the half-on-shell (HOS) $T$ matrix equation~\cite{NNInt}
\begin{equation}
T(k',k;k^{2}) = V_{\text{NN}}(k',k) + \frac{2}{\pi} \, \mathcal{P}
\int_{0}^{\infty} \frac{V_{\text{NN}}(k',p) \,
T(p,k;k^{2})}{k^{2}-p^{2}} \, p^{2} dp .
\label{fulltmat}
\end{equation}
In a RG treatment of two-body scattering, we impose a
cutoff $\Lambda$ on all the loop integrals in the $T$ matrix equation
and replace the bare $V_{\text{NN}}$ with an effective potential
$V_{\text{low k}}$. The effective potential is independent of the
scattering energy and is allowed to depend on $\Lambda$ so that
$T(k',k;k^{2})$ for $k,k'<\Lambda$ is preserved. Clearly this is 
sufficient to ensure that low energy observables, which are fully on-shell
quantities are independent of $\Lambda$. Explicitly, we have
\begin{equation}
T(k',k;k^{2})= V_{\text{low k}}(k',k) + \frac{2}{\pi} \, \mathcal{P}
\int_{0}^{\Lambda} \frac{V_{\text{low k}}(k',p) \,
T(p,k;k^{2})}{k^{2}-p^{2}} \, p^{2} dp .
\label{lowtmat}
\end{equation}
By construction, the low momentum HOS $T$ matrices
calculated from Eq. (\ref{fulltmat}) and Eq. (\ref{lowtmat}) are
identical. Consequently, the low momentum components of the low energy
scattering states are preserved. We denote the standing wave scattering
states of the effective theory by $| \chi_k \rangle$, which can be written 
in the standard fashion as
\begin{equation}
| \chi_k \rangle = | k \rangle + \frac{2}{\pi} \, \mathcal{P}
\int_{0}^{\Lambda} \, p^{2} dp \, \frac{1}{k^{2}-p^{2}} \, T(p,k;k^2)
\, | p \rangle .
\end{equation}
We therefore take $d T(k',k;k^{2}) /d \Lambda =0$ in Eq. (\ref{lowtmat})
to obtain a flow equation for $V_{\text{low k}}$ 
\begin{equation}
\int_{0}^{\Lambda} \frac{d V_{\text{low k}}(k',p)}{d \Lambda} \:
\chi_{k}(p) \: p^{2} dp = \frac{2}{\pi} \frac{V_{\text{low
k}}(k',\Lambda) \, T(\Lambda,k;k^{2})} {1 - (k/\Lambda)^{2}} .
\end{equation}
By writing $T(\Lambda,k;k^{2}) = \langle \Lambda | V_{\text{low
k}} | \chi_{k} \rangle$ and using the completeness of the
scattering states in the model space~\footnote{In the appendix,
we show that HOS T matrix preservation 
implies a non-hermitian $V_{\text{low k}}$.
Therefore, one must use the bi-orthogonal complement 
$\langle \widetilde{\chi}_{k} |$ in the completeness relation.
In the cutoff range of interest, the non-hermiticity 
for the nucleon-nucleon problem is generally small and can be
transformed away~\cite{KLR}.}, we obtain
\begin{align}
\frac{d}{d\Lambda} V_{\text{low k}}(k',p') & =
\frac{2}{\pi} \, V_{\text{low k}}(k',\Lambda)
\int_{0}^{\Lambda} \frac{\langle \Lambda | V_{\text{low k}} | \chi_{k}
\rangle \: \langle \widetilde{\chi}_{k} | p' \rangle}{1-(k/\Lambda)^{2}} \:
k^2 dk \nonumber \\[1mm]
& = \frac{2}{\pi} \, \Lambda^{2} \, V_{\text{low k}}(k',\Lambda)
\int_{0}^{\Lambda} V_{\text {low k}}(\Lambda,p) \:
\mathcal{G}(p,p';\Lambda^{2}) \: p^{2} dp .
\label{Gflow}
\end{align}
where $\mathcal{G}$ denotes the interacting Green's function in
the effective theory. Writing the interacting Green's function in
terms of the $T$ matrix in the low momentum sector
we obtain the exact RG equation
\begin{equation}
\frac{d}{d \Lambda} V_{\text{low k}}(k',k) = \frac{2}{\pi}
\frac{V_{\text{low k}}(k',\Lambda) \: T(\Lambda,k;\Lambda
^{2})}{1-(k / \Lambda)^{2}} .
\label{Tflow}
\end{equation}
The derivation of Eqs. (\ref{Gflow},\ref{Tflow}) is similar to the
work of Birse {\it et al.}~\cite{RG2body}, although they consider
the fully off-shell $T$ matrix and an energy dependent effective
potential $V_{\text{eff}}$. The RG equation of Birse {\it et al.}
is exact to second order in $V_{\text{eff}}$. It reads
\begin{equation}
\frac{d}{d \Lambda} V_{\text{eff}}(k',k;p^{2}) = \frac{2}{\pi}
\frac{V_{\text{eff}}(k',\Lambda;p^{2}) \:
V_{\text{eff}}(\Lambda,k;p^{2})}{1-(p/ \Lambda)^{2}} ,
\label{flowbirse}
\end{equation}
where $p^2$ denotes the scattering energy.

Our RG treatment is based on HOS $T$ matrix preservation, 
as this is the minimally off-shell quantity needed to 
preserve the low energy phase shifts as well as the low momentum
components of the wave functions, while allowing one to consider an
energy independent effective interaction. Our exact RG equation for energy
independent effective interactions, Eq.~(\ref{Tflow}), agrees with
the result of Birse {\it et al.}~\cite{RG2body} at the one-loop level.
As we shall see below, the starting point of energy independent effective
interaction methods is the above off-shell effective potential, and the 
energy dependence is subsequently traded for momentum dependence. 
This procedure is similar to using the equations of motion in
many-body physics to trade the energy dependence in favor of a more
complicated quasiparticle dispersion relation.

\section{RG from the KLR Folded Diagram Corrections}

In the previous section, we obtained the RG equation for $V_{\text{low
k}}$, Eq. (\ref{Gflow}), from $T$ matrix equivalence alone. We now
derive the same equation using the Kuo-Lee-Ratcliff (KLR)
folded diagram technique, originally designed for constructing
energy independent model space interactions in shell model
applications~\cite{KLR}. First, we define an energy dependent vertex
function called the $\widehat{Q}$ box, that is irreducible with respect to
cutting intermediate low momentum propagators. The $\widehat{Q}$
box resums the effects of the high momentum modes we wish to
integrate out of the theory,
\begin{equation}
\widehat{Q}(k',k;\omega) = V_{\text{NN}}(k',k) + \frac{2}{\pi} \, \mathcal{P}
\int_{\Lambda}^{\infty} \frac{V_{\text{NN}}(k',p) \, \widehat{Q}
(p,k;\omega)}{\omega-p^{2}} \, p^{2} dp .
\label{QVbare}
\end{equation}
It is the same energy dependent effective potential studied by
Weinberg in the context of chiral lagrangians~\cite{Weinberg} and by
Birse {\it et al.} for the purpose of an RG treatment of two-body
scattering~\cite{RG2body}. In terms of the $\widehat{Q}$ box the HOS
$T$ matrix reads
\begin{equation}
T(k',k;k^{2}) = \widehat{Q}(k',k;k^{2}) + \frac{2}{\pi} \,
\mathcal{P} \int_{0}^{\Lambda} \frac{\widehat{Q}(k',p;k^{2}) \,
T(p,k;k^{2})}{k^{2}-p^{2}} \, p^{2} dp .
\label{TQeq}
\end{equation}
The original literature develops KLR folded diagrams using
time-ordered perturbation theory for systems with degenerate
unperturbed model space spectra. In this framework, the folded
diagrams can be viewed as correction terms one must add, if one
factorizes the nested time-integrals one finds in time-ordered
perturbation theory for Dyson equations. For the
present work, it is much easier to understand the meaning of
folded diagrams in the time-independent framework. Iterating
Eq.~(\ref{TQeq}), it is clear that all $\widehat{Q}$ box vertices
are fully off-shell with the exception of the final $\widehat{Q}$
box in each term. The 
folded diagrams are correction terms one must add, if
all $\widehat{Q}$ boxes are evaluated right-side on-shell in the
Dyson equation. In order to see how this works, the second order
$\widehat{Q}$ box contribution in Eq.~(\ref{TQeq}) can be written as
\begin{equation}
\int_p \frac{\widehat{Q}(k',p;k^{2}) \,
\widehat{Q}(p,k;k^{2})}{k^{2}-p^{2}} = \int_p
\frac{\widehat{Q}(k',p;p^{2}) \, \widehat{Q}(p,k;k^{2})}{k^{2}-p^{2}}
\: + \: \mathcal{V}^{1-\text{fold}}(k',k) ,
\end{equation}
where $\mathcal{V}^{\text{1-fold}}$ denotes the ``one-fold
correction'' and $\int_{p} \equiv 2/\pi \int_0^\Lambda p^2 dp$
is shorthand for the model space principle value measure. With the
expression for $\mathcal{V}^{\text{1-fold}}$, we write the
KLR folded diagram series for $V_{\text{low k}}$ to one-fold, or
equivalently to second order in the $\widehat{Q}$ box as
\begin{multline}
V_{\text{low k}}(k',k) = \widehat{Q}(k',k;k^{2}) \\
+ \int_p
\frac{\widehat{Q}(k',p;k^{2}) - \widehat{Q}(k',p;p^{2})}{k^{2}-p^{2}}
\, \widehat{Q}(p,k;k^{2}) \: + \mathcal{O}(\widehat{Q}^3) .
\end{multline}
Analogously, the two-fold correction to the KLR $V_{\text{low k}}$ is
obtained by considering the $\mathcal{O}(\widehat{Q}^{3})$
contribution of Eq.~(\ref{TQeq}) and correcting for putting all
$\widehat{Q}$ boxes right-side on-shell. The resulting two-fold
correction is given by
\begin{multline}
\mathcal{V}^{2-\text{fold}}(k',k)=
\iint\limits_{p,p'}
\frac{\widehat{Q}(k',p';k^{2})-\widehat{Q}(k',p';p^{\prime2})}
{(k^{2}-p^{\prime2})(k^{2}-p^{2})} \, \widehat{Q}(p',p;k^{2}) \,
\widehat{Q}(p,k;k^{2})\\[1mm]
- \iint\limits_{p,p'}
\frac{\widehat{Q}(k',p';p^{2})-\widehat{Q}(k',p';p^{\prime2})}
{(p^{2}-p^{\prime2})(k^{2}-p^{2})} \, \widehat{Q}(p',p;p^{2}) \,
\widehat{Q}(p,k;k^{2}).
\end{multline}
Proceeding in a similar manner for the higher order terms in
the Dyson equation, one can write the KLR expression for
$V_{\text{low k}}$ as
\begin{equation}
V_{\text{low k}} \equiv \widehat{Q} + \mathcal{V}^{1-\text{fold}} +
\mathcal{V}^{2-\text{fold}} + \mathcal{V}^{3-\text{fold}} + \ldots \: .
\label{KLRV}
\end{equation}
Having motivated the KLR folded diagram series, we now derive a flow
equation that tells us how the KLR $V_{\text{low k}}$ changes as one
changes the scale $\Lambda$. The first step is to
notice that the $\widehat{Q}$ box, and hence the KLR series, obeys 
the semi-group composition law. We can therefore perform
the decimation to $\Lambda$ in one step using the bare
$V_{\text{NN}}$ as input, or equivalently we can integrate out recursively in
small steps; each subsequent step in the decimation uses the $V_{\text{low k}}$
of the previous step as input. Let us therefore start with the 
KLR $V_{\text{low k}}$ at a cutoff $\Lambda$, and ask how 
$V_{\text{low k}}$ changes as we integrate out the
states lying in the momentum shell $\Lambda - \delta\Lambda < k <
\Lambda$. Referring to the definition of the $\widehat{Q}$ box,
Eq.~(\ref{QVbare}), it is clear that each intermediate
loop integration is $\mathcal{O}(\delta\Lambda)$.
Therefore, we have for $\widehat{Q}^{\Lambda-\delta\Lambda}$
\begin{multline}
\widehat{Q}^{\Lambda - \delta\Lambda}(k',k;p^{2}) =
V^{\Lambda}_{\text{low k}}(k',k) \\
- \delta\Lambda \, \frac{2}{\pi} \, \frac{V^{\Lambda}_{\text{low
k}}(k',\Lambda)\, V^{\Lambda}_{\text{low
k}}(\Lambda,k)}{1-(p/\Lambda)^{2}} \, + \mathcal{O}(\delta\Lambda^{2}),
\label{Qinfdec}
\end{multline}
where the superscript denotes at which step the low momentum potential
and the $\widehat{Q}$ box are evaluated. Although we have given
explicit formulas only up to two folds, this will be sufficient to
derive the flow equation. The generalization to all
orders in $V_{\text{low k}}$ can be found in the appendix, where
energy independent effective interactions are related to the spectral
representation of the HOS $T$ matrix. We insert
the expression for the $\widehat{Q}^{\Lambda-\delta\Lambda}$ box,
Eq.~(\ref{Qinfdec}), into the KLR folded diagram series,
Eq.~(\ref{KLRV}), and find
\begin{multline}
\delta V_{\text{low k}}(k',k) = \delta\Lambda \, \frac{2}{\pi} \,
\frac{V^{\Lambda}_{\text{low k}}(k',\Lambda)}{1-(k/\Lambda)^{2}} \, \biggl\{ \,
V^{\Lambda}_{\text{low k}}(\Lambda, k) + \int_p
\frac{V^{\Lambda}_{\text{low k}}(\Lambda,p) \,
V^{\Lambda}_{\text{low k}}(p,k)}{\Lambda^{2}-p^{2}} \\[1mm]
+ \iint\limits_{p,p'} \frac{V^{\Lambda}_{\text{low k}}(\Lambda,p) \,
V^{\Lambda}_{\text{low k}}(p,p')
\, V^{\Lambda}_{\text{low k}}(p',\Lambda)}{(\Lambda^{2}-
p^{2})\, (\Lambda^{2}-p^{\prime2})} +
\mathcal{O}\bigl( (V^{\Lambda}_{\text{low k}})^4 \bigr) \, \biggr\} \, +
\mathcal{O}(\delta\Lambda^{2}) .
\end{multline}
Passing to the limit $\delta\Lambda \to 0$, we sum the flow equation
and obtain
\begin{equation}
\frac{d}{d\Lambda} V_{\text{low k}}(k',k) = \frac{2}{\pi} \, \frac{V_{\text{low
k}}(k',\Lambda) \, T(\Lambda,k;\Lambda^{2})}{1-(k/\Lambda)^{2}} .
\end{equation}
This proves that the KLR $V_{\text{low k}}$ obeys the same scaling
equation one obtains using RG arguments based on $T$ matrix
preservation alone. It does not come as a surprise, as it has been shown
diagrammatically that the KLR effective theory preserves the
low momentum HOS $T$ matrix of the input bare
potential~\cite{Vlowkdiag}. Nevertheless, we have here derived the same
flow equation in a manner that makes no reference
to $T$ matrix preservation. It may be regarded as an algebraic proof that
the folded diagram series preserves the HOS $T$ matrix.
This firmly establishes that the KLR folded diagram theory is equivalent to a
RG decimation, at least for the two-nucleon sector.

\section{RG from the LS Similarity Transformation}

It is easy to see how the folded diagram series is constructed.
The n-folded diagram is the correction terms one must add,
if one evaluates all soft-mode irreducible vertex functions right-side
on-shell in the term of $\mathcal{O}(\widehat{Q}^{n+1})$ of any Dyson
equation. As we have seen above, the expression for
the two-fold contribution is already rather cumbersome. The Lee-Suzuki
scheme is an iterative method that is often used in
the calculation of shell model effective interactions~\cite{LS}. It has been
shown to converge to the folded diagram series~\cite{KLR}. While the LS
method does not lend itself to a simple diagrammatic interpretation, 
we show how
the same flow equation evolves from the iteration scheme, as this provides a
check to our RG equation. The details of the LS method may be found
in~\cite{LS}. The defining equations of the LS iteration are given by
\begin{align}
H_{\text{low k}}^{(1)} & = PH_{0}P + \widehat{Q}(\omega) 
\arrowvert_{\omega=0} \\
H_{\text{low k}}^{(2)} & = \frac{1}{1-\widehat{Q}^{(1)}} \, H_{\text{low
k}}^{(1)} \\
H_{\text{low k}}^{(3)} & = \frac{1}{1-\widehat{Q}^{(1)}-\widehat{Q}^{(2)}
\, H_{\text{low k}}^{(2)}} \, H_{\text{low k}}^{(1)} \\
& \hspace{0.2cm} \vdots \nonumber \\
H_{\text{low k}}^{(n)} & = \frac{1}{1-\widehat{Q}^{(1)}-\widehat{Q}^{(2)}
\, H_{\text{low k}}^{(n-1)} - \ldots -\widehat{Q}^{(n-1)} \, H_{\text{low
k}}^{(2)} \, \cdots \, H_{\text{low
k}}^{(n-1)}} \, H_{\text{low k}}^{(1)} \\[1mm]
\widehat{Q}^{(n)} & \equiv \frac{1}{n!} \, \frac{d^{n}}{d\omega
^{n}} \widehat{Q}(\omega) \arrowvert_{\omega=0} .
\end{align}
As for the KLR folded diagram theory, we derive the flow equation for
$V_{\text{low k}}$ directly from the LS equations by
invoking the semi-group composition law and considering an
infinitesimal decimation from $\Lambda$ to $\Lambda -
\delta\Lambda$. Suppose we have solved the LS equations for
$V^{\Lambda}_{\text{low k}} = H^{\Lambda}_{\text {low k}} - P H_{0} P$
at some value of $\Lambda$, where $P$ denotes the projection operator
on the model space. Keeping terms to first order in $\delta\Lambda$, we find
\begin{align}
\widehat{Q}^{\Lambda-\delta\Lambda}(p',p;0) & =
V^{\Lambda}_{\text{low k}}(p',p) - \delta\Lambda \, \frac{2}{\pi} \,
V^{\Lambda}_{\text{low k}}(p',\Lambda) \, V^{\Lambda}_{\text{low k}}(\Lambda,p)
\\[1mm]
\widehat{Q}^{(n) \Lambda-\delta\Lambda}(p',p) & = - \delta\Lambda \,
\frac{2}{\pi} \, \frac{V^{\Lambda}_{\text{low k}}(p',\Lambda) \,
V^{\Lambda}_{\text{low k}}(\Lambda,p)}{\Lambda^{2n}} .
\end{align}
Substituting the above expressions into the defining LS equations and
keeping terms only to $\mathcal{O}(\delta\Lambda)$, one obtains
\begin{equation}
H^{\Lambda-\delta\Lambda}_{\text{low k}} \, (p',p) =
H^{\Lambda}_{\text{low k}}(p',p)
- \delta\Lambda \: V^{\Lambda}_{\text{low k}}(p',\Lambda) \int_k
V^{\Lambda}_{\text{low k}}(\Lambda,k) \: \mathcal{S}(k,p) ,
\end{equation}
where
\begin{equation}
\mathcal{S}(k,p) = \frac{1}{k^{2}} \: \delta(k-p) + \int_q
\frac{H^{\Lambda}_{\text{low k}}(k,q)}{\Lambda^{2}} \:
\mathcal{S}(q,p) .
\end{equation}
The expression for $\mathcal{S}$ can be summed to give the interacting
Green's function $\mathcal{S}(k,p) = \Lambda^{2} \,
\mathcal{G}(k,p;\Lambda^{2})$ of the effective theory.
In the limit $\delta\Lambda \to 0$, we thus obtain the
following RG equation directly from the Lee-Suzuki equations
\begin{align}
\frac{d}{d\Lambda}V_{\text{low k}}(p',p) & = \frac{2}{\pi} \, \Lambda^{2}
\: V_{\text{low k}} (p',\Lambda) \int_{0}^{\Lambda} V_{\text
{low k}}(\Lambda,k) \: \mathcal{G}(k,p;\Lambda^{2}) \: k^{2} \, dk \\[1mm]
& = \frac{2}{\pi} \frac{V_{\text{low k}}(p',\Lambda) \:
T(\Lambda,p;\Lambda ^{2})}{1-(p / \Lambda)^{2}} .
\end{align}
Thus, we obtain the same scaling equation from the LS and KLR effective
interaction methods as we obtain from RG arguments based on $T$ matrix
equivalence. Hence, it is clear that the traditional model space methods
of nuclear structure can be employed to perform RG decimations in the two
nucleon sector.

\section{Numerical Comparison of the Beta Function}

We have provided four different derivations of the same RG equation, 
which describes the scaling of the effective potential 
$V_{\text{low k}}$ with $\Lambda$ so that the
low energy observables are preserved by the effective theory. For 
completeness, we provide numerical confirmation of the RG equation
in this section. Starting from a model of the nuclear force such as the Paris,
Bonn-A, Argonne V-18, or Idaho-A potential, we have integrated out
the model dependent high momentum modes of the theory using the KLR/LS
methods to obtain an effective potential $V_{\text{low k}}$~\cite{Vlowk}.  
The results of Bogner {\it et al.} are striking~\cite{Vlowk}: the high
momentum modes of all studied potential models renormalize the low
momentum interaction to a unique $V_{\text{low k}}$ for cutoffs
$\Lambda \lesssim 2 \; \text{fm}^{-1}$. The differences resulting
from a chiral treatment of the 2$\pi$ exchange can be seen in the
offdiagonal matrix elements of $V_{\text{low k}}$, and flow studies of
the crucial s-wave matrix elements $V_{\text{low k}}(0,0)$ exhibit the
relevant scales (pions, deuteron, and higher order tensor
interactions) and for small model spaces are in excellent agreement
with the power divergence subtraction scheme of Kaplan, Savage, and
Wise~\cite{KSW}. For a discussion of all results see~\cite{Vlowk}.

\begin{figure}[b]
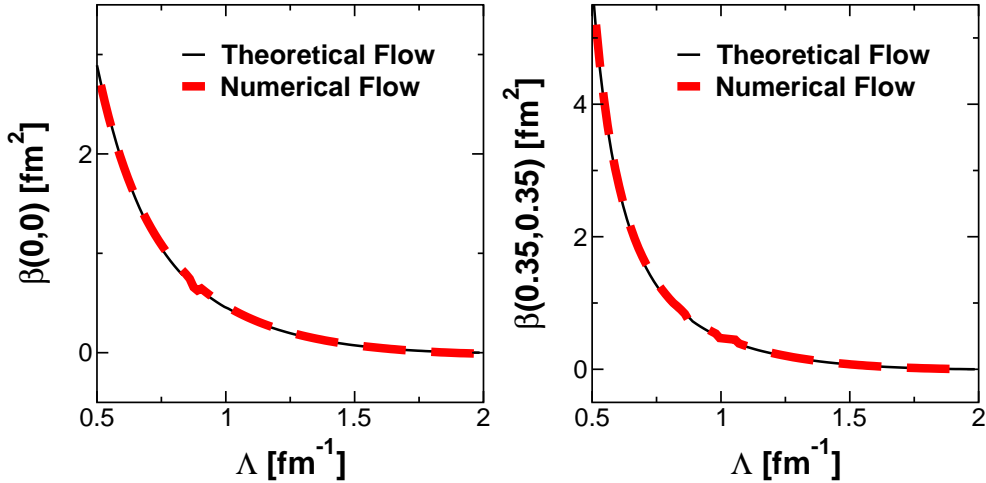

\includegraphics[scale=.35,clip=]{1s0flowzero.eps}
\includegraphics[scale=.35,clip=]{1s0flowsmall.eps}
\caption{Comparison of numerical and theoretical ``beta'' functions in
the $^{1}S_{0}$ partial wave calculated for the Paris potential.} 
\label{1S0flowdiag}
\end{figure}
 
Performing the calculation for a range of $\Lambda$ values,
we calculate the numerical derivative $d V_{\text{low k}} / d \Lambda$
and compare it with the theoretical expression, Eq.~(\ref{Tflow}). 
As in~\cite{Vlowk}, the RG decimation is carried out with the iterative
implementation of the LS method of Andreozzi~\cite{Andreozzi}.
Referring to Fig.~(\ref{1S0flowdiag}), it is evident that the
numerical derivative (numerical flow) is in excellent agreement 
with the exact ``beta'' function (theoretical flow)~\footnote{Technically 
speaking, this is not a true beta function as conventionally one
rescales all dimensional quantities in units of $\Lambda$. Our units
are such that $\hbar^{2}/m=1$.}, $\beta \equiv
\frac{d}{d\Lambda}V_{\text{low k}}$. Similar results are found for
offdiagonal elements as well as other partial waves. 

\section{Conclusions}

We have shown that model space effective interaction methods,
such as the Kuo-Lee-Ratcliff folded diagram series and the Lee-Suzuki
similarity transformation, scale with the model space boundary exactly
as one would expect from RG arguments. We have derived the exact beta 
function for the energy independent, low momentum effective theory of   
two-body scattering. Our theoretical beta function is
in perfect agreement with the numerical scaling behaviour of 
$V_{\text{low k}}$. This is a genuinely new result, as previous 
RG studies of two-body scattering consider energy dependent 
effective interactions~\cite{RG2body}.
 
In the RG formulation for effective theories, one can evolve the 
effective interaction according to the beta function given suitable
initial conditions; in our case, the bare interaction for very large
model spaces. Bogner {\it et al.} have performed such an RG decimation
in the two-nucleon sector starting from realistic nucleon-nucleon 
potentials such as the Paris, Bonn-A, Argonne V-18, and Idaho-A
interaction~\cite{Vlowk}. They found that the initially very 
different bare potentials renormalize to a unique $V_{\text{low k}}$ 
starting at a scale of $\Lambda \approx 2 \; \text{fm}^{-1}$. 
Moreover, the authors have used the unique $V_{\text{low k}}$ directly 
in shell model calculations without first calculating the Brueckner
$G$ matrix. They found very promising results using the same input 
$V_{\text{low k}}$ for two valence nuclei in different mass regions, 
and the results were nearly independent of the cutoff in the
neighborhood of $2 \; \text{fm}^{-1}$~\cite{O18}. Referring to 
Fig.~(\ref{1S0flowdiag}), these results are consistent with the 
observation that the beta function is nearly zero in this region.

For energy dependent effective interactions, Birse {\it et al.} have
studied the power counting imposed by means of a Wilson-Kadanoff
rescaling. Their results have been very powerful in predicting only
two well defined power-counting schemes for s-wave
scattering~\cite{RG2body} with short ranged forces. Their arguments 
can be extended to include long-range forces in a distorted wave 
basis~\cite{RG2bodylong}, with qualitatively similar conclusions. 
We hope to generalize and apply the methods of Birse {\it et al.} 
to obtain a well defined power counting scheme for the energy
independent $V_{\text{low k}}$. This is complicated since the ``beta''
function includes $V_{\text{low k}}$ to all orders.

In conclusion, the equivalence of model space effective interaction and RG
methods in the two-body sector is a first step towards reformulating
traditional effective interaction methods in a manner that fully
exploits their similarities to RG-EFT methods, particularily their
power to reliably estimate errors. The ultimate goal is to extend the
RG arguments for $V_{\text{low k}}$ to the shell model effective
interaction. At this point, it is restricted to two-particle correlations
in the nucleus.

\begin{ack}
This work is supported by the US-DOE grant DE-FG02-88ER40388.
\end{ack}

\appendix
\section{Appendix}

In this appendix, we provide the fourth and final derivation of the RG
equation, Eq.~(\ref{Tflow}), based on the spectral representation of the HOS
$T$ matrix. Moreover, we show that HOS $T$ matrix preservation necessarily
implies a non-hermitian energy independent $V_{\text{low k}}$. 
Without loss of generality, we employ a notation which suggests there are
no bound states in the channel under consideration; bound states are 
easily handled by extending the completeness relation to include the relevant
bound states in the summation.

We first prove that the energy independent $V_{\text{low k}}$ 
is necessarily non-hermitian, if it preserves the HOS $T$ matrix.
We begin by assuming the contrary, i.e. we assume the energy 
independent $V_{\text{low k}}$ is hermitian and preserves the 
HOS $T$ matrix. We have
\begin{equation}
V_{\text{low k}} | \chi_{p} \rangle = \widehat{Q}(p^{2}) | \chi_{p} \rangle .
\label{nec}
\end {equation} 
Multiplying from the right with $\langle \chi_p |$ and
integrating over $p$, the completeness of the model space eigenfunctions yields
\begin{equation}
V_{\text{low k}}(k',k) = \int_p \: \langle k' | \widehat{Q}(p^{2}) |
\chi_{p} \rangle \: \chi_{p}^{*}(k) .
\label{nec1}
\end {equation}
Expanding the $p^{2}$ dependence of the $\widehat{Q}$ box, one can
use the equations of motion $(H_{0}+\widehat{Q}(p^{2})) |\chi_p\rangle
= p^{2} |\chi_p\rangle$ to convert the $p^{2}$ dependence to
$k^{\prime 2}$
dependence. For example, the $\mathcal{O}(p^{2})$ dependence is
eliminated by letting $H_{0}$ act on $\langle k'|$ and using
Eq.~(\ref{nec}) to trade the extra $\widehat{Q}$ for $V_{\text{low
k}}$. It is clear from this simple example that the energy dependence
is asymmetrically converted to $k'$ dependence. 
Therefore, we conclude that the requirement of HOS $T$ matrix preservation
necessarily implies a non hermitian $V_{\text{low k}}$. 
Technically speaking, this means that the completeness relation for
the model space scattering states should be modified by replacing 
$\langle\chi_p|$ with the bi-orthogonal complement,
$\langle\widetilde{\chi}_p|$.

Having shown that $V_{\text{low k}}$ is non-hermitian, we now 
derive the RG equation from the spectral representation of the HOS
$T$ matrix.  
Inserting a complete set of low momentum plane waves between the
$\widehat{Q}$ box and the low energy scattering wave function in 
Eq.~(\ref{nec1}), we obtain 
\begin{equation}
V_{\text{lowk}}(k',k) = \iint\limits_{p,p^\prime}
\: \widehat{Q}(k',p^\prime;p^2) \: \chi_{p}(p^\prime) \:
\widetilde{\chi}_{p}^{*}(k). 
\label{Eavchi}
\end{equation}
In terms of $T$ matrices, this can be written as
\begin{equation}
V_{\text{low k}}(k^\prime,k) =  T(k^\prime,k;k^2) + \int_p \: T(k^\prime,p;p^2)
\: \frac{1}{p^2 - k^2} \: T(p,k;p^2) . 
\label{EavTT}
\end{equation}
This is just a rearrangement of the spectral representation for the 
$T$ matrix. Differentiating Eq.~(\ref{EavTT}) with respect to the cutoff 
we obtain the RG flow equation
\begin{align}
\frac{d V_{\text{low k}}}{d \Lambda}(k^\prime,k) = &
\, \frac{2}{\pi} \: V_{\text{low
k}} (k^\prime,\Lambda) \: \frac{\Lambda^2}{\Lambda^2 - k^2} \:
T(\Lambda,k;\Lambda^2) \nonumber \\
& + \frac{2}{\pi} \: \int_p \: V_{\text{low k}}(k^\prime,p) \:
\frac{1}{\Lambda^2 - p^2} \: T(p,\Lambda;\Lambda^2) \:
\frac{\Lambda^2}{\Lambda^2 - k^2} \:
T(\Lambda,k;\Lambda^2) \nonumber \\
& + \int_p \: T(k^\prime,p;p^2) \: \frac{1}{p^2-k^2} \:
\frac{d T(p,k;p^2)}{d \Lambda} . \label{apprg1}
\end{align}
Now we show that the latter two terms in Eq. (\ref{apprg1}) cancel
against each other. To this end, we write the left-side on-shell
$T$ matrix in terms of the right-side on-shell one:
\begin{equation}
T(p,k;p^2) = T(p,k;k^2) + \int_q \: T(p,q;q^2) \: \biggl\{
\frac{1}{p^2 - q^2} - \frac{1}{k^2 - q^2} \biggr\} \: T(q,k;q^2) .
\label{leftright}
\end{equation}
By differentiating with respect to the cutoff, Eq.~(\ref{leftright}) yields
\begin{multline}
\frac{d T(p,k;p^2)}{d \Lambda} = \frac{2}{\pi} \: T(p,\Lambda;\Lambda^2) \:
\biggl\{ \frac{\Lambda^2}{p^2 - \Lambda^2} - \frac{\Lambda^2}{k^2
- \Lambda^2}
\biggr\} \: T(\Lambda,k;\Lambda^2) \\
+ \int_q \: T(p,q;q^2) \: \biggl\{ \frac{1}{p^2 - q^2} -
\frac{1}{k^2 - q^2} \biggr\} \: \frac{d T(q,k;q^2)}{d \Lambda} .
\label{rshosrg}
\end{multline}
We insert the integral equation for the right-side on-shell $T$
matrix, Eq.~(\ref{lowtmat}), into the last term of flow equation,
Eq. (\ref{apprg1}). It results in the two terms
\begin{align}
& \int_p \: T(k^\prime,p;p^2) \: \frac{1}{p^2-k^2} \:
\frac{d T(p,k;p^2)}{d \Lambda}
= \int_p \: V_{\text{low k}}(k^\prime,p) \:
\frac{1}{p^2-k^2} \: \frac{d T(p,k;p^2)}{d \Lambda} \nonumber \\[1mm]
& \hspace{1cm} + \iint\limits_{p,q} \: V_{\text{low
k}}(k^\prime,q) \: \frac{1}{p^2 - q^2} \: T(q,p;p^2) \:
\frac{1}{p^2-k^2} \: \frac{d T(p,k;p^2)}{d \Lambda} .
\label{lastterm}
\end{align}
Finally, we combine Eq.~(\ref{lastterm}) with the integral equation
for $dT/d\Lambda$, Eq. (\ref{rshosrg}), inserted into the $V_{\text{low k}} \,
dT/d\Lambda$ term. Now the last two terms of the flow equation,
Eq. (\ref{apprg1}), read
\begin{align}
& \frac{2}{\pi} \, \int_p \: V_{\text{low k}}(k^\prime,p) \:
\frac{1}{\Lambda^2 - p^2} \: T(p,\Lambda;\Lambda^2) \:
\frac{\Lambda^2}{\Lambda^2 - k^2} \:
T(\Lambda,k;\Lambda^2) \nonumber \\[1mm]
& \; + \, \frac{2}{\pi} \, \int_p \: V_{\text{low k}}(k^\prime,p) \:
\frac{1}{p^2-k^2} \: T(p,\Lambda;\Lambda^2) \: \biggl\{
\frac{\Lambda^2}{p^2 - \Lambda^2} - \frac{\Lambda^2}{k^2 -
\Lambda^2}
\biggr\} \: T(\Lambda,k;\Lambda^2) \nonumber \\[1mm]
& \;\;\; + \iint\limits_{p,q} \: V_{\text{low k}}(k^\prime,p)
\: \frac{1}{p^2-k^2} \:  T(p,q;q^2) \: \biggl\{ \frac{1}{p^2 -
q^2} - \frac{1}{k^2 - q^2} \biggr\} \: \frac{d
T(q,k;q^2)}{d \Lambda} \nonumber \\[1mm]
& \;\;\;\; + \iint\limits_{p,q} \: V_{\text{low k}}(k^\prime,p)
\: \frac{1}{q^2 - p^2} \: T(p,q;q^2) \: \frac{1}{q^2-k^2} \:
\frac{d T(q,k;q^2)}{d \Lambda} ,
\end{align}
and we find the result that these terms cancel pairwise upon combining
the energy denominators, resulting in the familiar RG equation. Therefore,
we have provided the final derivation of the ``beta'' function for the two-body
scattering RG, Eq.~(\ref{Tflow}), based on the spectral representation of
the $T$ matrix. Moreover, we have shown that $V_{\text{low k}}$ is necessarily
non-hermitian.

\end{document}